\documentclass{ws-ijmpaMY}
\usepackage{hyperref}

\begin{document}

\markboth{A. A. Grib, Yu. V. Pavlov, O. F. Piattella}
{High energy processes in the vicinity of the Kerr's black hole horizon}

\title{\uppercase{High energy processes in the vicinity of the Kerr's
black hole horizon}}

\author{\footnotesize A. A. GRIB\footnote{
Present address:
Theoretical Physics and Astronomy Department, The Herzen  University,
48, Moika, St.\,Petersburg, 191186, Russia}}

\address{
A. Friedmann Laboratory for Theoretical Physics,\\
30/32 Griboedov can., St.\,Petersburg, 191023, Russia\\
andrei\_grib@mail.ru}

\author{YU. V. PAVLOV}

\address{
Institute of Mechanical Engineering, Russian Academy of Sciences,\\
61 Bolshoy, V.O., St.\,Petersburg, 199178, Russia\\
yuri.pavlov@mail.ru}

\author{O. F. PIATTELLA}
\address{
Departamento de F\'isica, Universidade Federal do Esp\'irito Santo,\\
avenida Ferrari 514, 29075-910 Vit\'oria, Esp\'irito Santo, Brazil,\\
INFN sezione di Milano, Via Celoria 16, 20133 Milano, Italy\\
oliver.piattella@gmail.com}

\maketitle

\begin{abstract}
    Two  particle collisions close to the horizon of the rotating
nonextremal black hole are analyzed.
    It is shown that high energy of the order of the Grand Unification scale
in the centre of mass of colliding particles can be obtained when there is
a multiple collision -- the particle from the accretion disc gets
the critical momentum in first collision with the other particle close
to the horizon and then there is a second collision of the critical
particle with the ordinary one.
    High energy occurs due to a great relative velocity of two particles
and a large Lorentz factor.
    The dependence of the relative velocity on the distance to horizon is
analyzed, the time of movement from the point in the accretion disc to
the point of scattering with large energy as well as the time of back
movement to the Earth are calculated.
    It is shown that they have reasonable order.

\end{abstract}

\ccode{PACS numbers: 04.70.-s, 04.70.Bw, 97.60.Lf}

\section{Introduction}

    There is much interest today to the high energy processes in the
ergosphere of the Kerr's rotating black hole as the model for
Active Galactic Nuclei (AGN).
    In Ref.~\refcite{GribPavlov2007AGN} some of the authors of this paper put
the hypothesis that due to Penrose process and scattering in the vicinity
of the horizon superheavy particles of dark matter due to large centre
of mass energy transfer  can become ordinary particles observed on the Earth
as ultra high energy cosmic rays (UHECR) by
the AUGER group.\cite{Auger07,AugerCor10}

    In Ref.~\refcite{BanadosSilkWest09} a resonance for the centre of mass (CM)
energy of two scattering particles close to the horizon of the extremal Kerr's
black hole was found.
    A pole for some special value of the angular momentum of the scattered
particle was found, showing that the centre of mass energy can go to infinity
on the horizon.
    Let us call this effect the BSW effect.
    In our
papers\cite{GribPavlov2010}\cdash\cite{GribPavlov2011b}
it was shown that the BSW effect can occur for the nonextremal black hole if
one takes into account the possibility of multiple scattering of the particle:
in the first scattering close to the horizon the particle gets the angular
momentum close to the critical one.
    In the second scattering close to the first one the particles due to
BSW effect occur to be in the region of high energy physics --- Grand
Unification or even Planckean physics.
    In Ref.~\refcite{Zaslavskii10c}
(see also Refs.~\refcite{Zaslavskii10,HaradaKimura10})
it was shown that the BSW effect can be
connected with the special behaviour of the Killing vector in ergosphere.
    In this paper we continue our analysis of this process
made in Refs.~\refcite{GribPavlov2010}--\refcite{GribPavlov2011b}.

    First we get a new formula of the dependence of the energy in the centre
of mass frame for two scattering particles  after the first scattering
on some distance to horizon.
    We show that the physical reason of the resonance effect is in the large
relative velocity of two particles being close to the velocity of light
leading to a big Lorentz factor.
    One of these particles has a critical angular momentum,
the other noncritical one.
    The noncritical particle has the velocity close to that of light,
the critical  one moving at some angle has a small velocity.
    This coincides with the point recently expressed by
O.\,Zaslavskii in Ref.~\refcite{Zaslavskii11}.

    Surprisingly this energy decreases (but surely being of the same order)
when going to horizon.
    This means that to get the high energy the second scattering can occur
close to the point of the first scattering and there is no need for
the coordinate time to be large for a particle to arrive to the region
of high CM energy.
    The decrease of the energy with the decrease of the distance to horizon
can be understood as caused by the growth of both velocities of two particles
close to horizon.

     Then we estimate the time needed for a particle to come from
the accretion disc of the AGN to the point of the first scattering.
    It occurs that this time can be of the order of the week!
    Then we evaluate the time needed for a high energy particle to leave
the ergosphere of the black hole to be observed on the Earth as UHECR.
    This time has reasonable order.

    The system of units $ G=c=1$ is used in the paper.

\section{The scattering energy in the centre of mass frame}

    The Kerr's metric of the rotating black hole in Boyer--Lindquist
coordinates has the form
    \begin{eqnarray}
d s^2 = d t^2 -
\frac{2 M r \, ( d t - a \sin^2 \! \theta\, d \varphi )^2}{r^2 + a^2 \cos^2
\! \theta } \hspace{33pt}
\nonumber \\
-\, (r^2 + a^2 \cos^2 \! \theta ) \left( \frac{d r^2}{\Delta} + d \theta^2 \right)
- (r^2 + a^2) \sin^2 \! \theta\, d \varphi^2, \ \
\label{Kerr}
\end{eqnarray}
    where
    \begin{equation} \label{Delta}
\Delta = r^2 - 2 M r + a^2,
\end{equation}
    $M$ is the mass of the black hole, $J=aM$ is angular momentum.
    The event horizon for the Kerr's black hole corresponds to the value
    \begin{equation}
r = r_H \equiv M + \sqrt{M^2 - a^2} .
\label{Hor}
\end{equation}
    The Cauchy horizon is
    \begin{equation}
r = r_C \equiv M - \sqrt{M^2 - a^2} . \label{HorCau}
\end{equation}

    For equatorial ($\theta=\pi/2$) geodesics in Kerr's metric~(\ref{Kerr}) one
obtains (see Ref.~\refcite{Chandrasekhar}, \S\,61)
    \begin{equation} \label{geodKerr1}
\frac{d t}{d \tau} = \frac{1}{\Delta} \left[ \left(
r^2 + a^2 + \frac{2 M a^2}{r} \right) \varepsilon - \frac{2 M a}{r} L \, \right],
\end{equation}
    \begin{equation}
\frac{d \varphi}{d \tau} = \frac{1}{\Delta} \left[ \frac{2 M a}{r}\,
\varepsilon + \left( 1 - \frac{2 M}{r} \right) L \, \right],
\label{geodKerr2}
\end{equation}
    \begin{equation} \label{geodKerr3}
\left( \frac{d r}{d \tau} \right)^2 = \varepsilon^2 +
\frac{2 M}{r^3} \, (a \varepsilon - L)^2 +
\frac{a^2 \varepsilon^2 - L^2}{r^2} - \frac{\Delta}{r^2}\, \kappa \,,
\end{equation}
    where
    $\kappa = 1 $ for timelike geodesics,
$\kappa = 0 $ for isotropic geodesics,
$\tau$ is the proper time for the moving particle with rest mass~$m \ne 0$,
$ L m = {\rm const} $ is the angular momentum of the particle relative
to the axis orthogonal to the plane of movement,
$\varepsilon={\rm const} $ is the specific energy:
the energy of particles in the gravitational field~(\ref{Kerr})
is  $\varepsilon m $.

    As it was shown in Ref.~\refcite{LL_II}, \S\,88,
the specific energy in the static gravitational field is  equal to
    \begin{equation} \label{bh1}
\varepsilon = \sqrt{ \frac{ g_{00} }{1 -\mathbf{v}^{2} } } ,
\end{equation}
    where $\mathbf v$ is the velocity of the particle measured by
the observer at rest at the point of the passing particle.
    So for the Schwarzschild black hole out of the event horizon
    \begin{equation} \label{bh1ddd}
\varepsilon = \sqrt{\frac{\displaystyle 1- \frac{r_g }{r} }
{\mathstrut 1 - \mathbf{v}^2 }} ,
\end{equation}
    where $r_g = 2 G M /c^2 $ is the gravitational radius,
$\mathbf v$ is the velocity of the particle measured by the local observer
at rest in the Schwarzschild coordinates.
    Expression~(\ref{bh1}) is still correct (see Ref.~\refcite{LL_II}, \S\,88)
in stationary field of the Kerr's metric if the velocity~$\mathbf v$ is
measured in proper time defined by the clock synchronized on
the trajectory of the particle.

    Let us find the energy $E_{\rm c.m.}$ in the centre of mass system
of two colliding particles with rest masses~$m_1$ and~$m_2$
in arbitrary gravitational field.
    It can be obtained from
    \begin{equation} \label{SCM}
\left( E_{\rm c.m.}, 0\,,0\,,0\, \right) =
m_1 u^i_{(1)} + m_2 u^i_{(2)}\,,
\end{equation}
    where $u^i=dx^i/ds$.
    Taking the squared~(\ref{SCM}) and due to $u^i u_i=1$ one obtains
    \begin{equation} \label{SCM2af}
\frac{E_{\rm c.m.}^{\,2}}{(m_1 + m_2)^2} = 1 + \frac{2 m_1 m_2}{(m_1+m_2)^2}
\left( u_{(1)}^i u_{(2) i} -1 \right).
\end{equation}
    This relation has maximal value for given $u_{(1)}, u_{(2)}$
and $ m_1 + m_2 $, if the particle masses are equal: $m_1 =m_2$.

    The scalar product does not depend on the choice of the coordinate frame
so~(\ref{SCM2af}) is valid in an arbitrary coordinate system and for arbitrary
gravitational field.

    Let us find the  expression of the energy in the centre of mass frame
through the relative velocity~$ v_{\rm rel}$ of particles at the moment
of collision.\cite{BanadosHassanainSilkWest10}
    In the reference frame of the first particle one has for
the components of 4-velocities  of particles at this moment
    \begin{equation} \label{Relsk01}
u_{(1)}^i = (1,0,0,0), \ \ \ \ \
u_{(2)}^i = \left( \frac{1 }{\sqrt{1- v_{\rm rel}^2}},\,
\frac{ \mathbf{v}_{\rm rel}}{\sqrt{1- v_{\rm rel}^2}} \right).
\end{equation}
    So
    \begin{equation} \label{Relsk02}
u_{(1)}^i u_{(2) i} = \frac{1}{\sqrt{1- v_{\rm rel}^2}}\,, \ \ \ \ \
v_{\rm rel} =
\sqrt{1- \frac{\mathstrut 1}{\left( u_{(1)}^i u_{(2) i} \right)^2}}\,.
\end{equation}
    These expressions evidently don't depend on the coordinate system.

    From~(\ref{SCM2af}) and~(\ref{Relsk02}) one obtains
    \begin{equation} \label{Relsk03}
\frac{E_{\rm c.m.}^{\,2}}{(m_1 + m_2)^2} = 1 + \frac{2 m_1 m_2}{(m_1+m_2)^2}
\left( \frac{1}{\sqrt{1- v_{\rm rel}^2}} -1 \right)
\end{equation}
    and the nonlimited growth of the collision energy in the centre of mass
frame occurs due to growth of the relative velocity to the velocity of
light.\cite{Zaslavskii11}

    For $v \ll c$ formula~(\ref{Relsk03}) gives a well known nonrelativistic
expression for the collision energy in the centre of mass frame
    \begin{equation} \label{Relsk04}
E_{\rm c.m.} \approx (m_1 + m_2) c^2 + \frac{m_1 m_2}{m_1+m_2}
\frac{v_{\rm rel}^2}{2} \,.
\end{equation}

    Let us find the expression for the collision energy of two particles
freely falling at the equatorial plane of the rotating black hole.
    We denote~$x=r/M$, \ $ A=a/M$, \ $ l_n=L_n/M$,
    \begin{equation} \label{DenKBHxhc}
x_H = 1 + \sqrt{1 - A^2}, \ \ \
x_C = 1 - \sqrt{1 - A^2},
\end{equation}
    \begin{equation} \label{Delxxhxc}
\Delta_x = x^2 - 2 x + A^2 = (x - x_H) (x- x_C) .
\end{equation}

    Using~(\ref{geodKerr1})--(\ref{geodKerr3}) one obtains:
    \begin{eqnarray}
u_{(1)}^i u_{(2) i} =
\frac{1}{x \Delta_x} \Biggl[
\varepsilon_1 \varepsilon_2 \left( x^3 + A^2(x+2) \right) -
2A \left( l_1 \varepsilon_2 +l_2 \varepsilon_1 \right) + l_1 l_2 (2-x)
\nonumber \\
-\, \sqrt{ 2 \varepsilon_1^2 x^2 + 2 (l_1 - \varepsilon_1  A )^2 \!- l_1^2 x
+ (\varepsilon_1^2 \!- 1 ) x \Delta_x }
\nonumber  \\
\times \, \sqrt{ 2 \varepsilon_2^2 x^2 +
2 (l_2 - \varepsilon_2  A )^2 \!- l_2^2 x
+ (\varepsilon_2^2 \!- 1 ) x \Delta_x }\, \Biggr].
\label{KerrL1L2}
\end{eqnarray}

    For $\varepsilon_1 = \varepsilon_2=1$
(\ref{KerrL1L2}) gives
    \begin{eqnarray}
u_{(1)} \cdot u_{(2)} = \frac{1}{x \Delta_x} \Biggl[
x^3 + A^2 (x+2) - 2 A ( l_1 + l_2 ) + l_1 l_2 (2-x) -
\nonumber \\
- \, \sqrt{ 2 x^2 + 2 (l_1 -  A )^2 \!- l_1^2 x }
\, \sqrt{ 2 x^2 + 2 (l_2 - A )^2 \!- l_2^2 x
}\, \Biggr].
\label{KerrL1L2e1e21}
\end{eqnarray}

    The value of $\Delta_x$ is going to zero on the event horizon and
as it is seen from~(\ref{KerrL1L2}) the scalar product of four
vectors $u_{(1)}^i u_{(2) i}$ and the collision energy of particles on
the horizon can be divergent depending on the behavior of the denominator
of the formula.
    To find the limit $r \to r_H$ for the black hole with a given angular
momentum~$A$ one must take in~(\ref{KerrL1L2}) $x = x_H + \alpha$
with $\alpha \to 0 $ and do calculations up to the order~$\alpha^2$.
    Taking into account $ A^2 = x_H x_C$, $x_H + x_C=2$, after resolution
of uncertainties in the limit $\alpha \to 0 $ one obtains
    \begin{eqnarray}
\frac{E_{\rm c.m.}(r \to r_H)}{m_1 + m_2} =
\left[ 1+ \frac{m_1 m_2}{(m_1+m_2)^2} \,
\frac{ (l_{1H} l_2 \!-l_{2H} l_1)^2 +
4 (l_{1H} \!-\! l_{2H} + l_2 \!-\! l_1)^2}
{4 (l_{1H}-l_1) (l_{2H} -l_2)} \right]^{1/2} \!\!,
\label{KerrLime1e2e2f}
\end{eqnarray}
    where
    \begin{equation} \label{KerrlH}
l_{nH} = \frac{2 \varepsilon_n x_H}{A} =  \frac{2 \varepsilon_n}{A}
\left( 1 + \sqrt{1-A^2} \, \right) =
\frac{\varepsilon_n}{M \Omega_H}\,,
\end{equation}
$\Omega_H = A/2 r_H$ is horizon angular velocity.\cite{Wald}
    In special case $ \varepsilon_1 = \varepsilon_2 $
(for example for nonrelativistic on infinity
particles $ \varepsilon_1 = \varepsilon_2 =1 $)
formula~(\ref{KerrLime1e2e2f}) can be written as
    \begin{eqnarray}
\frac{E_{\rm c.m.}(r \to r_H)}{m_1 + m_2} =
\sqrt{ 1+ \frac{m_1 m_2}{(m_1+m_2)^2}
\frac{(4+l_H^2)\, (l_1 - l_2)^2}{4 (l_H-l_1) (l_H -l_2)}}.
\label{KerrLime1e2e}
\end{eqnarray}

    For the Schwarzschild black hole $(A=0)$ the energy of collision in
the centre of mass frame is
    \begin{eqnarray}
\frac{E_{\rm c.m.}(r \to r_H)}{m_1 + m_2} =
\sqrt{ 1+ \frac{m_1 m_2}{(m_1+m_2)^2}
\frac{ (\varepsilon_1 l_2 - \varepsilon_2 l_1 )^2 +
4 (\varepsilon_1 - \varepsilon_2)^2}{4 \varepsilon_1 \varepsilon_2 }} \,.
\label{SchwLime1e2e}
\end{eqnarray}

    As it can be seen from~(\ref{KerrLime1e2e2f}) the collision energy of
particle in the centre of mass frame goes to infinity on the horizon if
the angular momentum of one of the freely falling particles has
the value~$l_{nH}$.
    Is  falling of the particle with such a value of the angular momentum
on the horizon possible?
    For the case of the free fall from infinity on the nonextremal $A<1$
rotating black hole it is impossible.
    This can be seen from the fact that the expression~(\ref{geodKerr3})
on the horizon is going to zero for $l=l_H$ but it's derivative
for $A<1$ is negative.

    The massive particle free falling in the black hole with dimensionless
angular momentum~$A$ being nonrelativistic at infinity ($\varepsilon = 1 $)
to achieve the horizon of the black hole must have angular momentum
from the interval
    \begin{equation} \label{geodKerr5}
- 2 \left( 1 + \sqrt{1+A}\, \right) =l_L \le l
\le l_R = 2 \left( 1 + \sqrt{1-A}\, \right).
\end{equation}
    Note that for the mentioned limiting values the right hand side of
the formula~(\ref{geodKerr3}) (for $\varepsilon = 1 $) is zero for
    \begin{equation} \label{geodKerrxR}
x_R = \frac{l_R^{\,2}}{4} = l_R - A = 2 \left( 1+ \sqrt{1-A}\, \right) - A ,
\end{equation}
    \begin{equation} \label{geodKerrxL}
x_L = \frac{l_L^{\,2}}{4} = A - l_L = 2 \left( 1+ \sqrt{1+A}\, \right) + A
\end{equation}
    for $ l=l_R$ and $l=l_L$ correspondingly.
    Obviously
    \begin{equation} \label{xCHRL}
x_C \le x_H \le x_R \le x_L \,.
\end{equation}

\begin{figure}[ht]
\centering
\includegraphics[height=55mm]{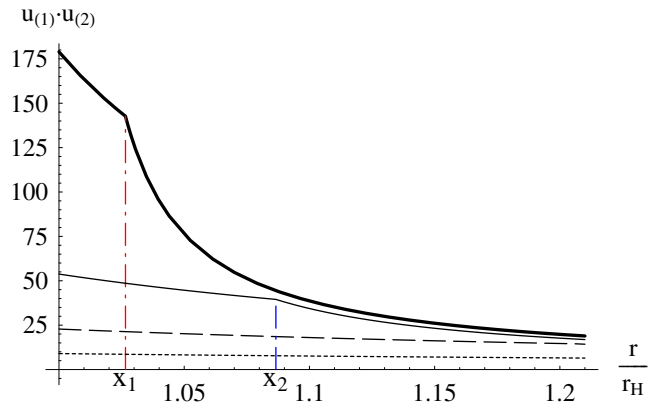}
\caption{The dependence of $u_{(1)}^i u_{(2) i}$ on the coordinate~$r$.}
\label{FigFA}
\end{figure}

     The dependence of the function $u_{(1)}^i u_{(2) i}$
for $\varepsilon_1=\varepsilon_2=1,\, l_1=l_R,\, l_2=l_L$ on the radial
coordinate is given on Fig.~\ref{FigFA}.
    The boldface solid line corresponds to~$A=0.998$,
the thin solid curve to~$A=0.98$, the dashed line to~$A=0.9$
and the dotted line describes non rotating black hole~($A=0$).
    The fractures on the solid lines, denoted by dash-dotted vertical
lines correspond to values $x=x_R(A) /x_H(A)$
(see~(\ref{DenKBHxhc}), (\ref{geodKerrxR})),
i.e. to zeroes of the expressions in roots in formula~(\ref{KerrL1L2e1e21}).

    The limiting values of the orbital momenta $l_R, l_L$ of
the nonrelativistic on infinity particles falling on the black hole
give the maximal value of the collision energy on the event horizon.
    Using the formula
    \begin{equation} \label{xLxR}
2 x^2 - l^2 x + 2 (A-l)^2 \Bigl|_{\, l= l_{R,\,L}} = 2 ( x - x_{R,\,L} )^2 ,
\end{equation}
    one obtains the explicit formula for the curves on Fig.~\ref{FigFA}
    \begin{equation} \label{FxLxR}
u_{(1)}^i u_{(2) i} = \left\{
\begin{array}{ll} 1+ \displaystyle
\frac{4 \bigl( \sqrt{\mathstrut x_H} + \sqrt{2}\, \bigr)^2}{x ( x - x_C)}
\,, & \ \ (x - x_L) ( x - x_R) \ge 0\,, \\[11pt]
1+ \displaystyle
\frac{4}{x - x_H}\,, & \ \ (x - x_L) ( x - x_R) \le 0\,.
\end{array}
\right.
\end{equation}
    From~(\ref{FxLxR}) and~(\ref{SCM2af})
one gets for $ l_1=l_R, \, l_2 =l_L $:
         \begin{equation} \label{ExLxR}
\frac{E_{\rm c.m.}^2}{(m_1 + m_2)^2} = \left\{
\begin{array}{ll} \displaystyle
1 + \frac{m_1 m_2}{(m_1+m_2)^2} \,
\frac{8 \bigl( \sqrt{\mathstrut x_H} + \sqrt{2}\, \bigr)^2}{x ( x - x_C)}
\,, & \ \ (x - x_L) ( x - x_R) \ge 0\,, \\[11pt]
\displaystyle
1 + \frac{m_1 m_2}{(m_1+m_2)^2} \,
\frac{8}{x - x_H}\,, & \ \ (x - x_L) ( x - x_R) \le 0 .
\end{array}
\right.
\end{equation}

    Formula~(\ref{FxLxR}) gives for the value on the horizon for $A<1$
the expression
    \begin{equation} \label{FxLxRHor}
u_{(1)}^i u_{(2) i} = 1 + \frac{4 \bigl( \sqrt{\mathstrut x_H} +
\sqrt{2}\, \bigr)^2}{x_H ( x_H - x_C)}\,.
\end{equation}
    That is why for the energy of collision on the horizon~(\ref{SCM2af})
one obtains the formula~(26) from Ref.~\refcite{GribPavlov2011b}.
    So even for values close to the extremal $A=1$ of the rotating black hole
$E_{\rm c.m.}^{\, \rm max}/ \sqrt{ m_1 m_2} $ can be not very large
as mentioned in Refs.~\refcite{BertiCardosoGPS09}, \refcite{JacobsonSotiriou10}
for the case $m_1=m_2$.
    So for $A_{\rm max} =0.998 $ considered as the maximal possible
dimensionless angular momentum of the astrophysical black holes
(see Ref.~\refcite{Thorne74}) one obtains
$ E_{\rm c.m.}^{\, \rm max} /\sqrt{m_1 m_2}\approx 18.97 $.

    Does it mean that in real processes of particle scattering in
the vicinity of the rotating nonextremal black holes the scattering energy
is limited so that no Grand Unification or even Planckean energies can
be obtained?
    Let us show that the answer is no!
    If one takes into account the possibility of multiple scattering so that
the particle falling from infinity on the black hole with some fixed
angular momentum changes its momentum in the result of interaction with
particles in the accreting disc and after this is again scattering close to
the horizon then the scattering energy can be unlimited.

    From~(\ref{geodKerr3}) one can obtain the permitted interval in~$r$ for
particles with $ \varepsilon = 1 $ and angular momentum $l = l_H - \delta $.
    To do this one must put the left hand side of~(\ref{geodKerr3})
to zero and find the root.
    In the second order in~$\delta$ close to the horizon one obtains
             \begin{equation} \label{xd}
l=l_H -\delta \ \ \Rightarrow  \ \ \
x_\delta = \frac{l^2 - \sqrt{ l^{\mathstrut 4} -16(l-A)^2 }}{4}
\approx x_H + \frac{\delta^2 x_C^2}{4 x_H \sqrt{1-A^{\mathstrut 2}} }
\end{equation}
    Note that smallness of the value of $x_\delta - x_H $ does not mean
the smallness of the ``physical distance'' from~$r_\delta$ to horizon
(see Ref.~\refcite{LL_II}, \S\,84).

\begin{figure}[ht]
\centering
\includegraphics[height=38mm]{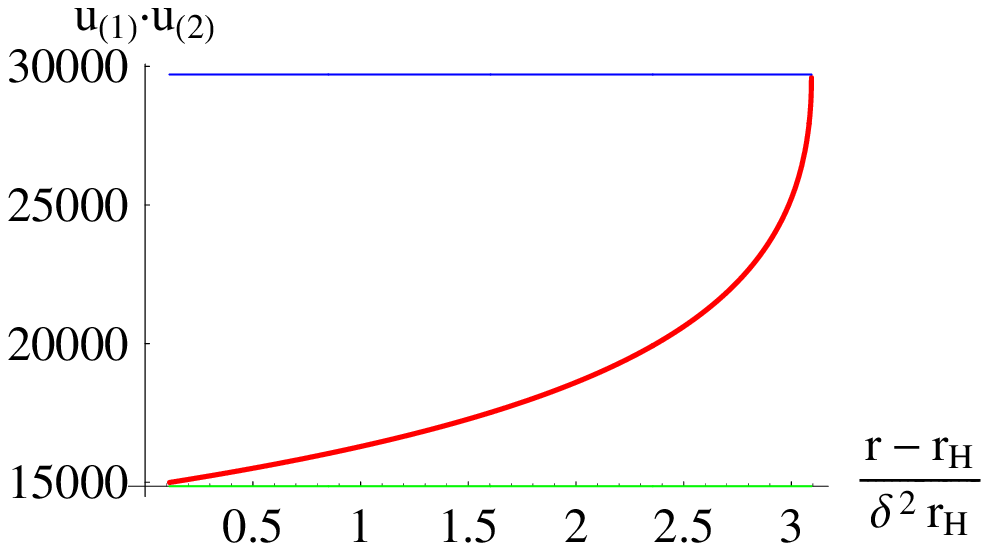}
\includegraphics[height=38mm]{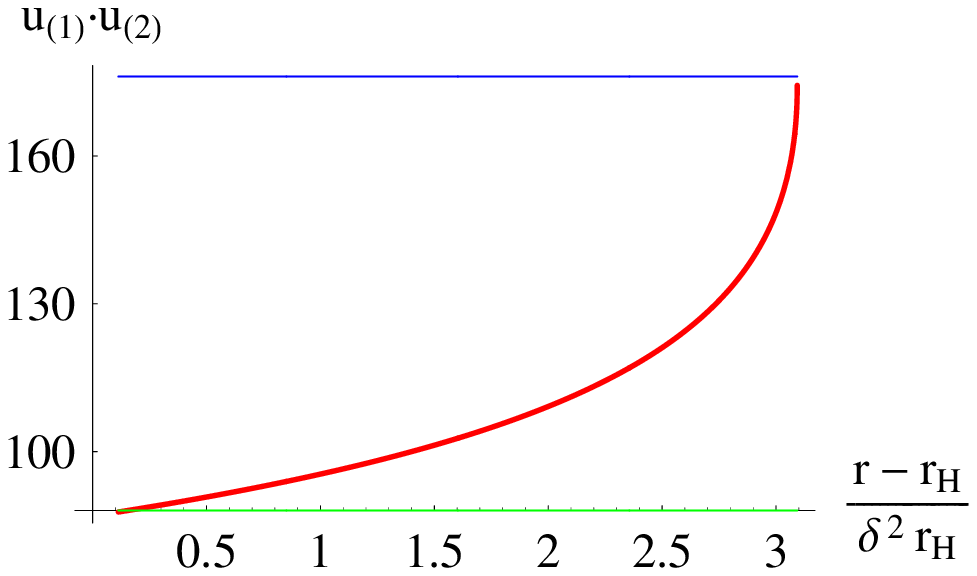}
\caption{The dependence of $u_{(1)}^i u_{(2) i}$ on the coordinate~$r$
for $A=0.998, \
\delta = 5 \cdot 10^{-4},\ l_2 = l_L$ on the left and $l_2=l_R$ on the right.}
\label{FigLR}
\end{figure}

    Values of function $u_{(1)}^i u_{(2) i}$ after first scattering
in points $x_\delta$ and on the horizon are
         \begin{equation} \label{FxdH}
\varepsilon_1=\varepsilon_2=1,\ \ l_1 = l_H-\delta \ \ \Rightarrow
\ \ \ u_{(1)}^i u_{(2) i} \, \approx \,
 \left\{
\begin{array}{ll} \displaystyle
2\,\frac{l_H - l_2}{\delta x_C} \,, & \ \ x= x_\delta , \\[11pt]
\displaystyle
\frac{l_H - l_2}{\delta x_C} \,, & \ \ x= x_H .
\end{array}
\right.
\end{equation}
    Therefore the function $u_{(1)}^i u_{(2) i}$ and hereby the energy of
collisions decrease near the horizon!
    The dependence of $u_{(1)}^i u_{(2) i}$ on the coordinate~$r$ is
shown on Fig.~\ref{FigLR}.

    The left picture shows that the collision energy can be very large
immediately after obtaining the angular momentum $ l_H - \delta$.
    But, for $A_{\rm max} =0.998 $, \ $ l_H - l_R \approx 0.04$ and so the
collision energy for particles with angular momentum $l_R$ is not large.
    This means that from the decrease of the collision energy with fixed
value of angular momentum does not follow the extremely large energy of
particles in the centre of mass frame needed to get the value $l_H - \delta$
in the intermediate collision.

    The decrease of the collision energy in the centre of mass frame of
the free falling particle and the particle with critical angular
momentum~$l=l_H-\delta$ in their movement to horizon is explained by
the decrease of the relative velocity of these particles when going from
the point~$x_\delta$.
    Due to the definition~(\ref{xd}) the radial velocity of the particle
with the angular momentum~$l=l_H-\delta$ is equal to zero:
$ d r / d \tau =0$ and $ d r / d t =0$.
     But the other noncritical particle colliding with the critical one
has large value of the radial velocity.
    Angular velocities $ d \varphi / d t $ for both particles go to
the same value $\Omega_H$.
    So large relative velocity of two particles occurs due to a ``stop''
of the critical particle in radial direction.
    But after this both particles increase their radial velocities
$ d r / d \tau $ and the relative velocity is decreasing --- the critical
particle is ``running down'' the free falling one.
    From~(\ref{Relsk02}), (\ref{FxdH}) one obtains for the relative velocity
of the particles colliding at the point~$x_\delta$ and on the horizon
             \begin{equation} \label{xdrel}
x=x_\delta \ \ \Rightarrow  \ \ \ v_{\rm rel} -1 =
\frac{\delta^2 x_C^2}{8 \left( l_H - l_2 \right)^2}\,,
\end{equation}
             \begin{equation} \label{xHrel}
x=x_H \ \ \Rightarrow  \ \ \ v_{\rm rel} -1 =
\frac{\delta^2 x_C^2}{2 \left( l_H - l_2 \right)^2}\,.
\end{equation}

    So the physical reason of the unlimited great energy of the collision
in the centre of mass frame of the particles falling in the black hole is
the increasing of the relative velocity of particles at the moment of
collision to the velocity of light.
    So one can expect very large energy of collision in the case when one
of the particles due to multiple intermediate collisions in the accretion
disc strongly diminishes its energy so that its velocity becomes small near
the horizon.
    Really from~(\ref{KerrLime1e2e2f}) it is easy to obtain that
    \begin{equation} \label{xHe1malo}
\frac{E_{\rm c.m.}}{\sqrt{m_1 m_2}} \sim
\sqrt{\frac{l_{2H} - l_2}{l_{1H} -l_1} } \to \infty\,,
\ \ \ \varepsilon_1 , l_1 \to 0 \,.
\end{equation}

    From the same considerations one can conclude that for the case of
nonrotating Schwarzschild black hole the collision energy of the free
falling particle with the particle at rest close to horizon also is
great and unlimited.
    Using~(\ref{bh1}), (\ref{SCM2af}), (\ref{KerrL1L2}) for $A=0$ and
the particle at rest in the point with radial coordinate~$r_0$
(so $l_1=0$, $\varepsilon_1 = \sqrt{1-r_g/r_0}$, $d r_1/d \tau=0$)
for the energy of its collision with the particle with $\varepsilon_2, l_2$
one obtains in the centre of mass frame
     \begin{equation} \label{ESchw}
E_{\rm c.m.}^{\,2} = m_1^2 + m_2^2 + 2 m_1 m_2 \varepsilon_2
\sqrt{ \frac{r_0}{r_0- r_g}},
\end{equation}
    which evidently is growing infinitely for $r_0 \to r_g$.

    If one particle in the point with radial coordinate~$r_0$
has $d r /d \tau=0$ but $l_1 \ne 0$, then from
(\ref{geodKerr3}), (\ref{SCM2af}), (\ref{KerrL1L2}) one has
     \begin{equation} \label{ESchwl}
E_{\rm c.m.}^{\,2} = m_1^2 + m_2^2 + 2 m_1 m_2 \left[ \varepsilon_2
\sqrt{ \frac{l_1^2 + x_0^2}{(x_0-2) x_0}} -\frac{l_1 l_2}{x_0^2} \right],
\end{equation}
    which also is growing infinitely for $x_0 \to x_H=2$.

    Note that for particles nonrelativistic on infinity with $m_1=m_2=m$,
freely falling on the Schwarzschild black hole the limiting energy of
collisions is only $2 \sqrt{5} m$ (see Ref.~\refcite{Baushev09}).

    Now let us show that for any fixed specific energy~$\varepsilon$
of the free falling particle in Kerr black hole with~$A<1$
there are no circular orbits for $r \to r_H$.

    Define the effective potential through the right hand side
of~(\ref{geodKerr3})
    \begin{equation} \label{Leff}
V_{\rm eff} = -\frac{1}{2} \left[\varepsilon^2 +
\frac{2 M}{r^3} \, (a \varepsilon - L)^2 +
\frac{a^2 \varepsilon^2 - L^2}{r^2} - \frac{\Delta}{r^2}\, \kappa \right].
\end{equation}
    Then
    \begin{equation} \label{LeffUR}
\frac{1}{2} \left( \frac{d r}{d \tau} \right)^{\!2} + V_{\rm eff}=0, \ \ \ \
\frac{d^2 r}{d \tau^2} = - \frac{d V_{\rm eff}}{d r}
\end{equation}
    and it is necessary for the existence of the circular orbit to have
    \begin{equation} \label{LeffCucl}
V_{\rm eff}=0, \ \ \ \ \frac{d V_{\rm eff}}{d r} =0\,.
\end{equation}
    Putting the effective potential to zero one has
    \begin{equation} \label{Leffl}
l = \frac{1}{2-x} \left( 2 A \varepsilon \pm \sqrt{
\Delta_x \left( \varepsilon^{\mathstrut 2} x^2 + \kappa x (2 - x)
\right) }\right).
\end{equation}
    The condition of the movement ``forward in time'' $d t / d \tau >0 $
(see~(\ref{geodKerr1})) for $x<2$ corresponds to taking the minus sign
in~(\ref{Leffl}).
    The corresponding value of~$l$ for $x \to x_H$ is going to~$l_H$
(see~(\ref{KerrlH})).
    For these values of~$l$ on horizon from~(\ref{Leff}) one has
    \begin{equation} \label{Lefflim}
A<1, \ \ x=x_H, \ \ l=l_H \ \ \Rightarrow \ \ \frac{d V_{\rm eff}}{d x} =
\frac{\sqrt{1-A^2}}{A^2}
\left( \varepsilon^2 + \kappa \frac{x_C}{x_H} \right) > 0 \,.
\end{equation}
    So for $A<1$ there are no circular orbits with $r \to r_H$
for all free falling particles including $l$ close to $l_H$.

    In conclusion of this part note that
the probability of the collision of the relativistic particle
with the particle at rest close to the Schwarzschild horizon is very small.
    So this is the main difference with the situation when
the BSW resonance occurs.
    This can be seen from the evaluation of the interval of the proper time
of falling from the point~$r_0$ where $dr/ d \tau =0$ to horizon.
    Let us do necessary calculations.
    \begin{equation} \label{tau}
\frac{d r}{ d \tau}\Bigl|_{r_0} = 0 \ \ \Rightarrow \ \
r \approx r_0 + \frac{\Delta \tau^2}{2} \frac{d^2 r}{d \tau^2}=
r_0 - \frac{\Delta \tau^2}{2} \frac{d V_{\rm eff}}{d r} \,.
\end{equation}
    So the proper time of falling of the particle to horizon is
    \begin{equation} \label{tau2}
\Delta \tau \approx M \sqrt{ \frac{2(x_0 - x_H)}
{\displaystyle \frac{d V_{\rm eff}}{d x}}}\,.
\end{equation}
    For the Schwarzschild black hole one obtains from~(\ref{Leff})
    \begin{equation} \label{tau3}
\Delta \tau \approx 2 M \sqrt{ \frac{2(x_0 - x_H)}
{\displaystyle 1+ \frac{l^2}{4}}}\,, \ \ x_0 \to x_H .
\end{equation}
    For the Kerr black hole taking $l$ close to $l_H$
from~(\ref{Lefflim}), (\ref{tau2}) one obtains
    \begin{equation} \label{tau4}
\Delta \tau \approx M A \sqrt{ \frac{2(x_0 - x_H)}
{\displaystyle \sqrt{1-A^2} \left( \varepsilon^2 + \frac{x_C}{x_H}
\right)}}\,, \ \ x_0 \to x_H ,
\end{equation}
    which evidently is much larger than~(\ref{tau3}) for $A\to 1$.

\section{Estimate of the time of the fall before collision leading to
the large energy.}

    In Refs.~\refcite{GribPavlov2010}--\refcite{GribPavlov2010c}
it was shown by us that in order to get the unboundedly growing energy
for the extremal case one must
have the time interval (as coordinate as proper time) from the beginning
of the falling inside the black hole to the moment of collision also growing
infinitely.
    Quantitative estimations have been given for a case of extremely rotating
black hole $A=1$.
    The time of movement before collision near horizon
with a given value of the energy~$E$ in the centre of mass frame
    \begin{equation} \label{telH3}
\Delta t \sim \frac{E^2}{m_1 m_2} \,
\frac{2 M \varepsilon_1}{ (2 \varepsilon_2 - l_2)
\sqrt{ 3 \varepsilon_1^2- 1 }
\,( 2 \varepsilon_1 - \sqrt{ 3 \varepsilon_1^2 - 1 } \,) } \,.
\end{equation}
    For $\varepsilon_1 = \varepsilon_2 = 1$ one gets
    \begin{equation} \label{telH4}
\Delta t \sim \frac{E^2}{m_1 m_2} \,
\frac{M }{ (2-l_2) (\sqrt{2} - 1 )} \approx \frac{1.2 \cdot 10^{-5}}{2-l_2}\,
\frac{M}{M_\odot}\, \frac{E^2}{m_1 m_2} \, {\rm s},
\end{equation}
    where $M_\odot$ is the mass of the Sun.
    So to have the collision of two protons with the energy of
the order of the Grand Unification one must wait for the black hole of
the star mass (and $l_2=0$) the time $\sim 10^{24}$\,s,
which is larger than the age of the Universe $\approx 10^{18}$\,s.
    However for the collision with the energy $10^3$ larger than
 that of the LHC one must wait only~$\approx 10^{8}$\,s.

    Here we consider the case $A<1$.

    From Eqs.~(\ref{geodKerr1}), (\ref{geodKerr3}) one gets
    \begin{equation} \label{telHe}
\left| \frac{dt}{dx}\right| =
\frac{ M \sqrt{x} \left((x^3 + A^2 x + 2 A^2) \varepsilon - 2 A l \right)}
{\Delta_x \sqrt{ 2 \varepsilon^{\mathstrut 2} x^2 - l^2 x +
2 (A \varepsilon - l)^2 + (\varepsilon^2 - 1) \Delta_x}} \,.
\end{equation}
    For $A<1$ from~(\ref{Delxxhxc}), (\ref{telHe}) one gets that
the value of the time interval measured by clock of the distant observer
necessary to achieve the horizon is logarithmically divergent.
    Remind that for the extremal black hole and the critical value of the
angular momentum of the falling particle this interval is divergent
as $ 1/(r-r_H)$.
    Let us give the estimates for the case $\varepsilon=1,\,l=A$, when the
expression for the time interval can be obtained in elementary functions.
    Taking the integral of~(\ref{telHe}) on the interval $(x_0,\, x_f)$
one obtains
    \begin{eqnarray}
\Delta t = M \Biggl\{
\sqrt{2 x - A^2} \left( 2 + \frac{x+A^2}{3} \right)
+ \frac{2}{x_H -x_C} \times \nonumber \\[2pt]
\Biggl[ x_C \log \frac{\left( x_C + \sqrt{2 x - A^2}\, \right)^{\!2}}
{2 (x-x_C)} - x_H \log \frac{\left( x_H + \sqrt{2 x - A^2}\, \right)^{\!2}}
{2 (x-x_H)} \Biggr] \Biggr\} \Biggr|_{\,x_0}^{\,x_f}.
\label{telHeInt}
\end{eqnarray}
    So for $\varepsilon=1,\,l=A$ one has
    \begin{equation} \label{telHelt}
\Delta t \sim - \frac{2 M x_H}{x_H - x_C} \log (x_f - x_H) \,, \ \ \
x_f \to x_H .
\end{equation}
    Note that the formula~(\ref{telHelt}) is valid for arbitrary~$\varepsilon$
and $l< l_H=2 \varepsilon x_H/A $,\, $A<1$, which follows from~(\ref{telHe}).

    From Eqs.~(\ref{SCM2af}), (\ref{xd}), (\ref{FxdH})
 it is easy to obtain for collisions of two particles with
$\varepsilon_1=\varepsilon_2 =1, \, l_1=l_H -\delta $
close to horizon at the point $x_\delta$
    \begin{equation} \label{tt2}
\frac{1}{x_\delta - x_H} \approx \frac{x_H(x_H-x_C)}{4 (l_H - l_2)^2}\,
\frac{E_{\rm c.m.}^{\,4}}{m_1^2 m_2^2}\,.
\end{equation}
    In this case from~(\ref{telHelt}) one obtains
    \begin{equation} \label{tt3}
\Delta t \sim \frac{8 M x_H}{x_H-x_C} \, \log \frac{E_{\rm c.m.}}
{\sqrt{m_1 m_2}}\,.
\end{equation}
    So for
    \begin{equation} \label{tt4}
A=0.998, \ \ \ \
\Delta t \sim 3.2 \cdot 10^{-4} \frac{M}{M_\odot} \,
\log \frac{E_{\rm c.m.}}{\sqrt{m_1 m_2}} \, {\rm s}.
\end{equation}
    Taking the value of the Grand Unification energy
$E_{\rm c.m.}/\sqrt{m_1 m_2}=10^{14}$ and the mass of
the black hole $10^8 M_\odot$ typical for Active Nuclei of galaxies
one obtains $\Delta t \sim 1.1 \cdot 10^6$\,s, i.e. of the order of 12.7 days.
    So in case of the nonextremal rotating black hole the mechanism of
the intermediate collision to get the additional angular momentum with
the following collision with other relativistic particle leading to large
collision energy proposed by us in Ref.~\refcite{GribPavlov2010}
needs reasonable time much smaller than that for the extremal case.

    One can ask about the time of back movement of the particle after
collision with very high energy from the vicinity of horizon to the Earth.
    Due to reversibility of equations of motion in time it is easy to see
that this time is equal to the sum of the same 12.7 days to accretion disc
and some 10--100 megaparsec --- the distance of the AGN from the Earth.

\section*{Acknowledgments}

    One of the authors A.A.G. is indebted to CAPES for financial support of
the work on the first part of the paper and to the UFES, Brazil,
for hospitality, other  author O.F.P. is thankful to CNPq for
partial financial support of his work.


\end{document}